\pdfoutput=1
\documentclass[10.5pt,letterpaper,fleqn, twocolumn]{article}

\usepackage{graphicx}
\usepackage{makeidx}
\usepackage{crop}
\usepackage{amstext,amsthm}
\usepackage{amsmath,amssymb}
\usepackage{bm}
\usepackage{natbib}
\usepackage{units}
\usepackage{tabularx}
\usepackage{booktabs}
\usepackage{afterpage}

\newenvironment{minimatrix}{\left[\begin{large}\begin{smallmatrix}}{\end{smallmatrix}\end{large}\right]}

\usepackage{multicol}
\usepackage{pslatex}
\usepackage[strict]{changepage}
\newcommand{\be}{\begin{equation}}
\newcommand{\ee}{\end{equation}}
\newcommand{\bea}{\begin{eqnarray}}
\newcommand{\eea}{\end{eqnarray}}
\newcommand{\bne}{\begin{equation*}}
\newcommand{\ene}{\end{equation*}}
\newcommand{\bi}{\begin{itemize}}
\newcommand{\ei}{\end{itemize}}

\newcommand{\bbm}{\begin{bmatrix}}
\newcommand{\ebm}{\end{bmatrix}}

\newcommand{\mr}{\mathrm}

\usepackage[top=0.75in, bottom=0.5in, left=0.83in, right=0.83in]{geometry}
\usepackage[hang,splitrule]{footmisc}



\makeatletter
\newenvironment{tablehere}
{\def\@captype{table}}
{}
\newenvironment{figurehere}
{\def\@captype{figure}}
{}
\makeatother

\usepackage{titlesec}
\usepackage{subcaption}
\usepackage{indentfirst}

\newenvironment{myabstract}
{
	\vspace{0.2in}
	\parindent=0cm \textit{Abstract}
	\parindent=0.5cm \hangindent=0.5cm \linebreak
}

\newenvironment{mykeywords}
{
	\vspace{0.2in}
	\parindent=0cm \textit{Keywords}
	
	\parindent=0.5cm
}

\titleformat{\section}
{\normalfont\bfseries}
{}{0.0pt}{\parindent=0cm}[\parindent=0.5cm]

\titlespacing*{\section}
{0pt}{1.5\baselineskip}{0.5\baselineskip}

\titleformat{\subsection}
{\normalfont\itshape}
{}{0.0pt}{\parindent=0cm}[\parindent=0.5cm]

\titlespacing*{\subsection}
{0pt}{1.0\baselineskip}{0.5\baselineskip}

\begin{document}
\twocolumn[{%

\phantom\\
\vspace{0.5in}

\iftrue
\vspace{-0.8in}
\begin{center}
	\textit{Published in Foundations of Computer Aided Process Operations / Chemical Process Control, FOCAPO (2023)}\\
\end{center}
\vspace{0.35in}
\fi

\begin{center}
\Large{\textbf{Data-Driven Model Reduction and Nonlinear Model Predictive Control of an Air Separation Unit by Applied Koopman Theory}}\\
\end{center}
\vspace{0.2in}

\begin{center}
Jan C.~Schulze $^{\mr{a}}$, Danimir T.~Doncevic $^{\mr{b}}$, Nils Erwes $^{\mr{a}}$ 
and Alexander Mitsos $^{\mr{c,a,b,}}$\footnotemark\\

\vspace{0.10in}

$^{\mr{a}}$ Process Systems Engineering (AVT.SVT), RWTH Aachen University, 52074 Aachen, Germany

\vspace{0.10in}
$^{\mr{b}}$ Institute of Energy and Climate Research: Energy Systems Engineering (IEK-10),\\ Forschungszentrum J\"ulich GmbH, 52425 J\"ulich, Germany

\vspace{0.10in}
$^{\mr{c}}$ JARA-CSD, 52056 Aachen, Germany

\end{center}

\begin{myabstract}
Achieving real-time capability is an essential prerequisite for the industrial implementation of nonlinear model predictive control (NMPC).
Data-driven model reduction offers a way to obtain low-order control models from complex digital twins.
In particular, data-driven approaches require little expert knowledge of the particular process and its model,
and provide reduced models of a 
well-defined generic
structure.
Herein, we apply our recently proposed data-driven reduction strategy based on Koopman theory [Schulze et al.~(2022), Comput.~Chem.~Eng.]~to generate a low-order control model of an air separation unit (ASU).
The reduced Koopman model combines autoencoders and linear latent dynamics 
and is constructed using machine learning.
Further, we present an NMPC implementation that uses derivative computation tailored to the fixed block structure of reduced Koopman models.
Our reduction approach with tailored NMPC implementation enables real-time NMPC of an ASU at an average CPU time decrease by \unit[98]{\%}.
\end{myabstract}

\begin{mykeywords}
Data-driven model reduction, 
Autoencoder, 
Koopman theory, 
Nonlinear model predictive control, 
Air separation unit
\end{mykeywords}
\vspace{0.2in}
}]

\footnotetext[1]{\parindent=0cm \small{Corresponding author. Email: {\tt amitsos@alum.mit.edu}.}}

\section{Introduction}
Computationally tractable models are a main requirement for real-time NMPC \citep{Marquardt.2002}.
Data-driven non-intrusive model reduction comprises 
a class of model-free methods
for producing low-order representations of high-order dynamical systems from data,
e.g., \cite{Antoulas.2017}.
Similar to classical model reduction approaches \citep{Marquardt.2002},
these data-driven methods project a high-order system from the full state space to a lower dimensional subspace, reducing the system to its dominant latent patterns.
However, applying data-driven approaches shifts the reduction efforts from system-specific expert knowledge for reduced modeling towards data-based system identification.
Notably, data-driven approaches allow for automated
reduction frameworks as well as exploitation
of the generic reduced model structure in optimization.

In recent years, a variety of data-driven frameworks for model reduction of dynamical systems
has been presented in the literature, e.g., the Loewner framework \citep{Antoulas.2017}, 
lift-and-learn methods \citep{Qian.2020}, and
autoencoder network structures \citep{Watter.2015}. 
In particular, Koopman theory \citep{Mezic.2005}
has attracted considerable attention.
Model reduction using Koopman theory
as well as the related dynamic mode decomposition \citep{Schmid.2010},
build on a lift-and-project concept and aim to construct linear representations of nonlinear dynamics through (nonlinear) coordinate transformation.
Applied Koopman theory has a system-theoretic foundation and naturally combines simple dynamic forms with data-driven identification of coordinate transformations, e.g., through Kernel methods \citep{Williams.2015c}, 
deep learning \citep{Lusch.2018},
or sparse regression techniques \citep{Brunton.2016b}.

While Koopman theory was originally developed for autonomous systems,
several extensions to systems with exogenous inputs have been introduced.
These works derive linear \citep{Proctor.2016, Korda.2018} 
as well as bilinear \citep{Surana.2016}
Koopman control models. 
Recently, we proposed a Wiener-type Koopman form \citep{Schulze.2022a}.
More specifically, we developed a Koopman-based deep learning framework that combines autoencoders networks with linear latent dynamic blocks.
Further, we demonstrated that such Wiener-type Koopman models can offer a favorable compromise between linear dynamics and nonlinear modeling.
Related works using autoencoders and latent dynamics have also been presented independent of Koopman theory, e.g., \citet{Watter.2015, Masti.2021}.
\newpage

Despite the potential of Koopman-based NMPC for process control, applications have been limited to single unit operations like a reactor \citep{Narasingam.2019} and a distillation column \citep{Schulze.2022b}. 
Herein, we aim to demonstrate that applied Koopman theory enables real-time NMPC of complex processes, while requiring little expert knowledge for the model reduction procedure.
We use our deep learning framework to train Wiener-type Koopman models on simulation data of a detailed digital process twin,
whose solution computation requirements are prohibitive for online applications.
As a further extension of our previous work \citep{Schulze.2022b}, we present an NMPC implementation that is tailored to the reduced Koopman structure, thereby realizing 
an additional speed-up in online optimization.
Finally, we apply the resulting Koopman NMPC framework to an ASU case study.
Our numerical results demonstrate the real-time capability and tracking performance of Koopman NMPC in load change scenarios.

\section{Koopman-based model reduction framework}
Koopman theory postulates the global linearization of nonlinear autonomous dynamics by means of nonlinear coordinate lifting to a high (generally infinite) dimensional coordinate space \citep{Mezic.2005}.
For practical application of Koopman theory,
finite truncation of this lifting is sought,
often using data-driven methods, e.g., 
\cite{Williams.2015c,Lusch.2018}.
We consider the class of asymptotically stable input-affine systems:
\begin{subequations}
	\label{eqn:observer}
	\begin{flalign}
	\dot{\bm x}(t) &= \bm f (\bm x(t)) + \sum_{i=1}^{n_u} \bm g_i(\bm x(t)) u_i(t) \,,\\
	\bm y(t) &= \bm h(\bm x(t)) \,, \label{eqn:outputs}
	\end{flalign}
\end{subequations}\noindent
where $\bm{x}(t) \in \mathbb{R}^{n_x}$  
are the differential states, 
$\bm y(t) \in \mathbb{R}^{n_y}$ are the outputs,
$\bm{u}(t) \in \mathbb{R}^{n_u}$ 
are external inputs, 
and 
$\bm f: \mathbb{R}^{n_x} \rightarrow \mathbb{R}^{n_x}$, 
$\bm g_i: \mathbb{R}^{n_x} \rightarrow \mathbb{R}^{n_x}$,
$\bm h: \mathbb{R}^{n_x} \rightarrow \mathbb{R}^{n_y}$
are continuously differentiable. 
For this system class, we have derived a Koopman representation of Wiener form \citep{Schulze.2022a, Schulze.2022b}, which uses linear time-invariant dynamics (LTI) in time-discrete form (zeroth-order hold),
sandwiched by nonlinear coordinate transformations:
\begin{subequations}
	\label{eqn:final}
	\begin{align}
	\bm z_{k+1} &= A \bm z_k + B \bm u_k  \,, \label{eqn:koopman_linear} \\
	\left[\begin{large}\begin{smallmatrix}\bm x_{k} \\ \bm y_{k}
	\end{smallmatrix}\end{large}\right]
	&= \bm T^{\dagger}(\bm z_{k})\,, \label{eqn:koopman_decoding}
	\\
	\bm z_0 &= \bm T(\bm x_0) \,.
	\end{align}
\end{subequations}
Herein, 
$\bm z_k \in \mathbb{R}^{n_z}$ 
are the Koopman coordinates,
$k$ is the time index,
$\bm T: \mathbb{R}^{n_x} \rightarrow \mathbb{R}^{n_z}$ (encoding) 
represents the nonlinear transformation to the Koopman coordinates and provides initial conditions, and
$\bm T^{\dagger}: \mathbb{R}^{n_z} \rightarrow \mathbb{R}^{n_x + n_y}$ (decoding) maps the Koopman coordinates to the original states and outputs.
$A$, $B$ are constant matrices.
Since we target model reduction, we focus on $n_z \ll n_x$.
Based on sampled process data, we aim to directly identify the discrete-time reduced form, which relieves the need for numerical integration in dynamic optimization.
Further, note that the dynamics in $z$ are linear instead of nonlinear, constituting a crucial simplification step for the computation of the Jacobian.

\subsection{Deep learning implementation}
We employ the Wiener-type Koopman form, Eq.~(\ref{eqn:final}), for data-driven model reduction.
To this end, we adopt our deep learning strategy from \citep{Schulze.2022a,Schulze.2022b},
where we train reduced models on sampled data from numerical simulations of the full-order model (e.g., a digital twin).
Specifically, we use artificial neural networks to learn suitable mappings
$\bm T$ and $\bm T^{\dagger}$.
To simplify the training, 
we modify
the mapping $\bm z=\bm T(\bm x)$ to $\bm z=\bm T(\bm x, \bm y)$, allowing us to train autoencoder networks.

We compute the training loss $\mathcal{C}$ as the sum of mean squared error (MSE) terms for single and multi-time-step predictions: 
\begin{equation}
\begin{split}
\mathcal{C} \;=\;\; \frac{1}{s-1}& \sum_{k=0}^{s-1} \left\| 
\begin{minimatrix}\bm x_{k+1}\\\bm y_{k+1}\end{minimatrix} 
- \bm T^{\dagger}(\bm z_{k+1}(k) 
) \right\|_{\mathrm{MSE}} \\
+\, \frac{1}{s-1}& \sum_{k=0}^{s-1} 
\left\| 
\begin{minimatrix}\bm x_{k+1}\\\bm y_{k+1}\end{minimatrix} 
- \bm T^{\dagger}(\bm z_{k+1}(0)
) \right\|_{\mathrm{MSE}} \,,\\
\end{split}
\end{equation}\noindent
where:
\begin{equation*}
\phantom{xxxx}
\bm z_{k+1}(j) :=
\begin{cases}
\bm z_{k+1} = A \bm z_{k} + B \bm u_{k} \,, \; k=j,j+1,...\\
\bm z_j = \bm T(\bm x_j, \bm y_j) \,,
\end{cases}
\end{equation*}\noindent
and $s$ is the number of snapshots per trajectory.
The dimension $n_z$ and number of network layers and respective neurons are hyperparameters. 
If system knowledge is available, a (block) diagonality of $A$ can be prespecified which enforces learning Koopman eigenfunctions and reduces the number of trainable parameters.
In contrast to our previous work,
we do not formulate an additional autoencoder loss term.
Thereby, we promote direct feedthrough of inputs to states and outputs, which improves 
the reduction procedure with respect to fast modes.
Case-specific details on the data sampling and model training procedures are provided within the control study.

\begin{figurehere}
    \vspace{1ex}
    \hspace{-1ex}
	\resizebox{0.95\columnwidth}{!}{\includegraphics{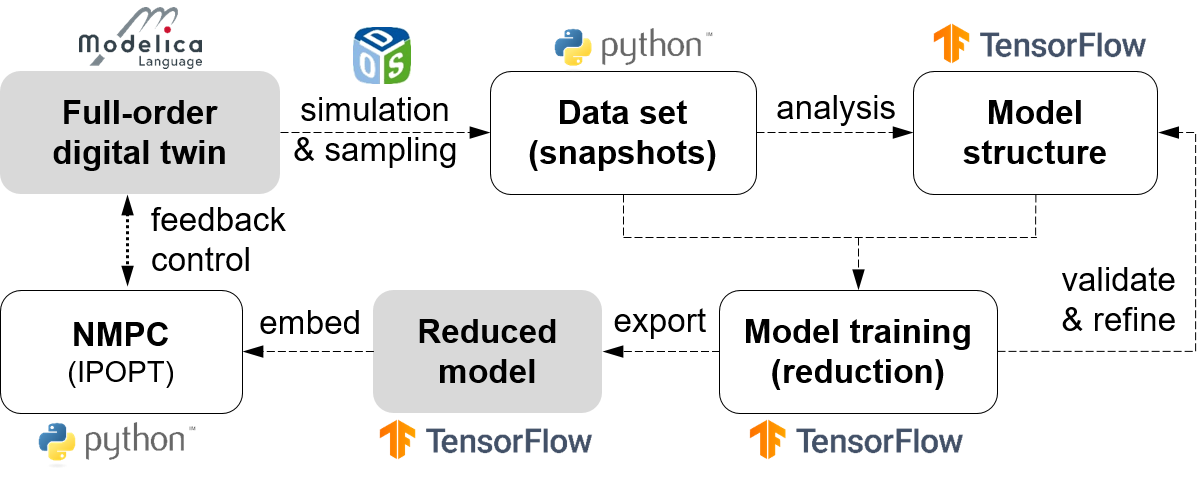}}
	\vspace{-2ex}
	\caption{\label{fig:workflow} Model reduction and control workflow.}
	\vspace{3ex}
\end{figurehere}

Fig.~\ref{fig:workflow} depicts the model reduction workflow implemented in Python 3.9 and TensorFlow 2.5.
We conduct the simulation experiments using our open-source dynamic optimization software DyOS with integrator NIXE \citep{CaspariDyOS}.
However, other simulation environments may be used as well.
The preliminary model structure is selected based on analysis of the simulation data set and refined iteratively by the user if necessary.
After training, the reduced model is used to control the real process (or as herein, its digital twin).
\vspace{-4ex}
\section{Koopman NMPC}
We employ the reduced Koopman models for NMPC of chemical processes.
While Wiener-type Koopman models promise a higher accuracy than linear models \citep{Schulze.2022a}, their nonlinear type requires nonlinear optimization.
However, the sequential Wiener architecture can be exploited by a tailored implementation of the NMPC.
Consider the following optimal control problem solved by the NMPC:
\begin{subequations}
	\label{eqn:nmpc}
	\begin{flalign}
	\min_{\bm u, \bm z} & \sum_{k=1}^{N_c} 
	\ell_k(\bm T^{\dagger}(\bm z_{k+1}))
	\label{eqn:nmpccost}\\ 
	\hspace{0ex}\mathrm{s.t.}\hspace{5ex}
	&\bm z_{k+1} = A \bm z_k + B \bm u_k  \,,
	\label{eqn:nmpc_koopman1}\\
	&\bm c(\bm T^{\dagger}(\bm z_{k+1})) \leq \bm 0\,, \label{eqn:nmpc_constraint} \\
	&\bm u_{k} \in \mathcal{U}\,,\\
	&\bm z_0 = \bm T(\bm x_0, \bm y_0) \,, \label{eqn:initial}\\
	&k=0,1,... , N_c - 1
	\,,
	\end{flalign}
\end{subequations}\noindent
where $N_c$ is the control horizon and
Eq.~(\ref{eqn:nmpccost}) describes the cost function with stage cost $\ell_k$.
Eq.~(\ref{eqn:nmpc_koopman1}) are the linear dynamics, 
Eq.~(\ref{eqn:nmpc_constraint}) represents nonlinear path constraints on states and outputs,
and $\mathcal{U}$ is the admissible set of controls.
Eq.~(\ref{eqn:initial}) provides process feedback and is evaluated prior to solving the problem.
Due to the sequential model structure, the decoding can be directly inserted into the cost function and constraints.
Consequently, the control problem is condensed to the variables $\bm u$ and $\bm z$.
Further, the sparsity of the Jacobian is known beforehand due to the well-defined model structure.
These attributes enable a computationally efficient and reusable Koopman NMPC framework.

We implement Koopman NMPC in Python 3.9 by combining the NLP solver IPOPT \citep{IPOPT} and automatic differentiation using TensorFlow.
Therein, we tailor the derivative computation to the block-diagonal sparse structure of the Jacobian 
and compute the non-zero entries using automatic differentiation.
We cache the gradient tape to avoid expensive online re-computation.
We use piecewise constant control moves (zeroth-order hold) and warm-start consecutive optimizations.
In contrast to, e.g., \cite{Masti.2021}, we exploit the model structure in the optimization rather than embedding the full model.
Fig.~\ref{fig:graph} visualizes how the underlying graph structure of the model and control problem predetermines the Jacobian. 
Neighboring $\bm z$ nodes are coupled through LTI dynamics, Eq.~(\ref{eqn:koopman_linear}), resulting in non-zero constant Jacobian blocks.
In addition, decoding branches for cost and constraints are locally attached, Eq.~(\ref{eqn:koopman_decoding}), with non-constant Jacobian blocks. 

\begin{figurehere}
    \vspace{2ex}
	\resizebox{0.85\columnwidth}{!}{\includegraphics{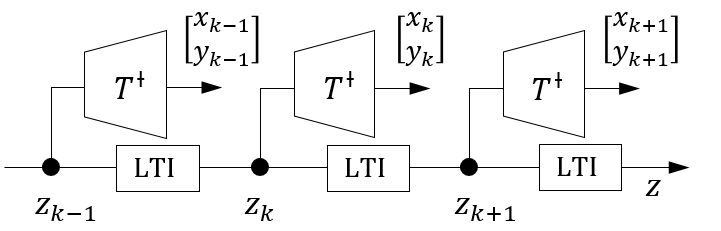}}
	\vspace{-1ex}
	\caption{\label{fig:graph} Graph structure of Koopman model prediction.}
\end{figurehere}

\section{Case study: NMPC of an air separation unit}
Cryogenic air separation units (ASUs) produce industrial gases such as nitrogen and oxygen. 
Due to their high electric energy demand, ASUs are considered as a promising candidate for load flexible operation and demand side management \citep{Pattison.2016b}.
NMPC is a promising operating strategy for realizing major load changes on a frequent basis \citep{Chen.2009}.
However, especially the detailed modeling of distillation columns and heat exchangers, e.g., by means of tray-to-tray balancing and finite volume discretization, respectively, results in process models that are too complex for NMPC \citep{CaspariWave}.
Thus, different model reduction approaches for individual process units have been investigated, e.g., \cite{Chen.2009, Schafer.2019b, CaspariWave, Schulze.2021}.
While these physics-motivated approaches preserve the flowsheet structure of the dynamic model,
their application requires additional modeling efforts and expert knowledge.
In contrast, data-driven reduction is particularly suitable if the flowsheet structure does not need to be preserved and the size and complexity of the input-output data sets is thus limited.

\begin{figurehere}
    \vspace{1ex}
	\resizebox{1.\columnwidth}{!}{\includegraphics{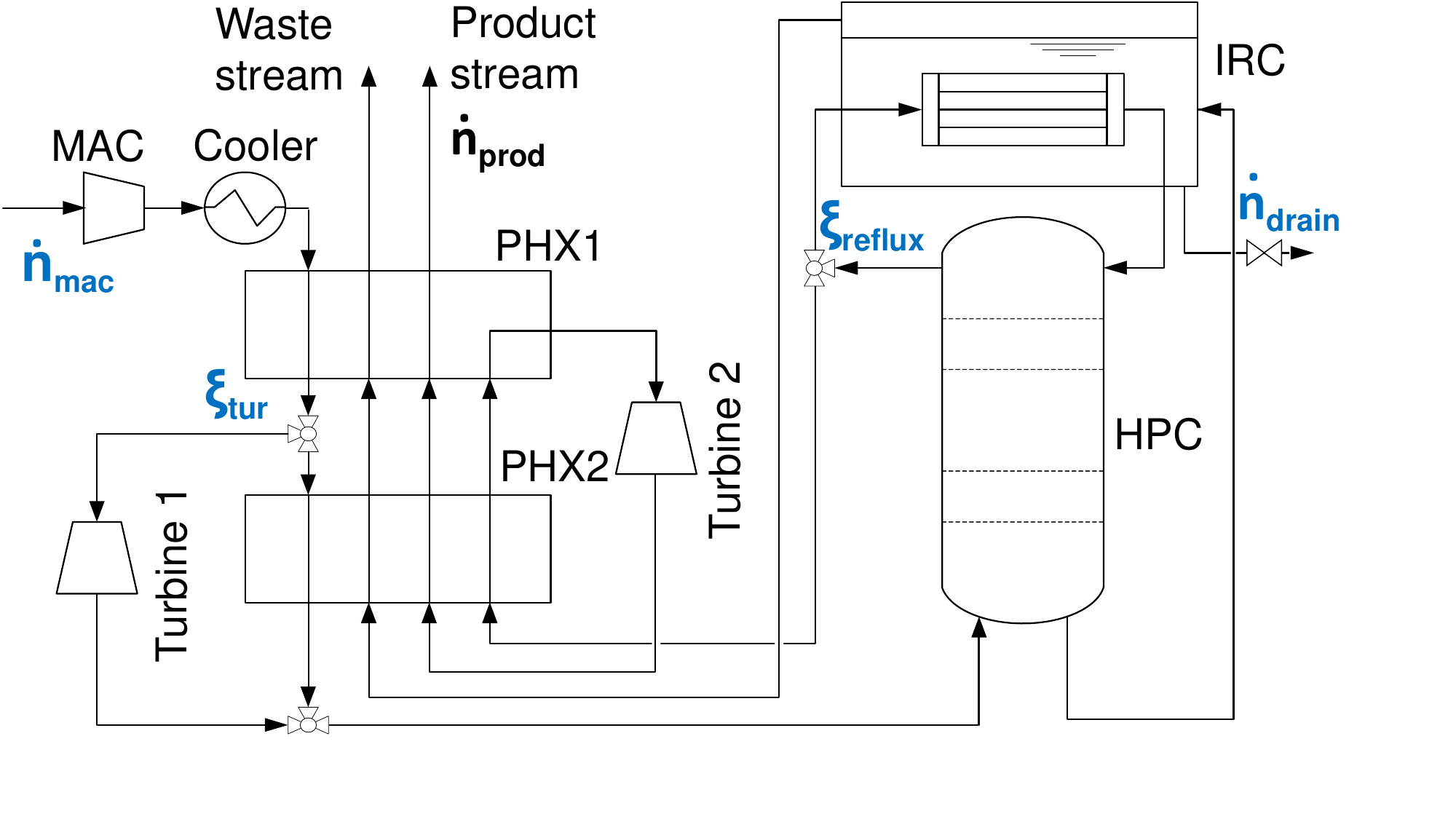}}
	\vspace{-6ex}
	\caption{\label{fig:asu} Air separation unit. Manipulated variables in blue.}
	\vspace{3ex}
\end{figurehere}

\begin{figure*}
\centering
	\includegraphics[width=0.93\linewidth]{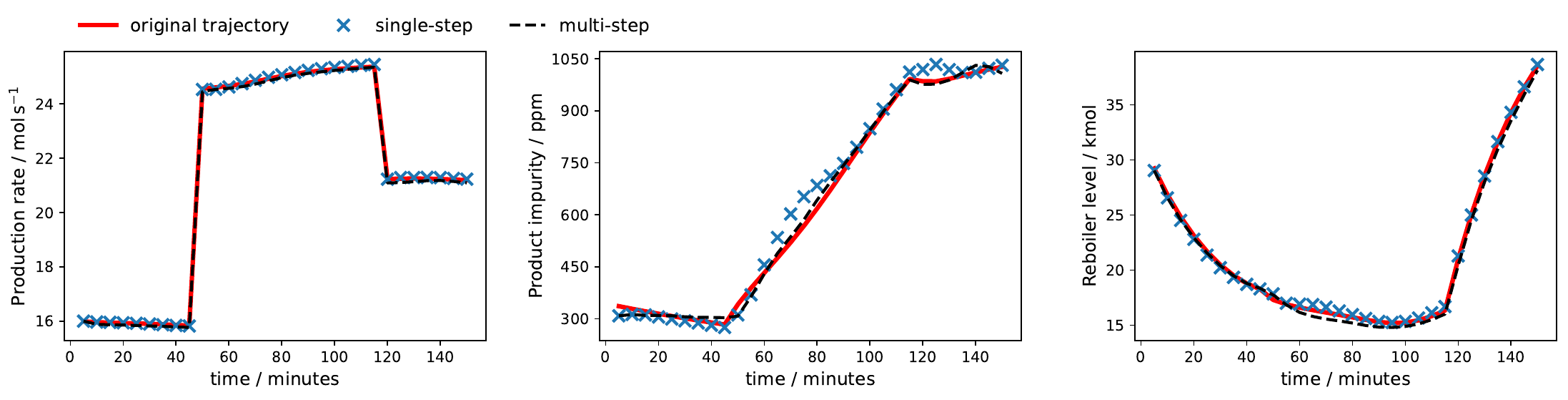}
	\vspace{-2ex}
	\caption{\label{fig:validation} Independent open-loop step test of the reduced Koopman model.}
	\vspace{-1ex}
\end{figure*}

We consider the ASU depicted in Fig.~\ref{fig:asu}, which serves as a classic literature problem \citep{Pattison.2016b}.
The process consists of the main air compressor (MAC), 
multi-stream heat exchangers (PHX),
turbines, 
high-pressure distillation column (HPC) 
and integrated reboiler condenser unit (IRC).
The nitrogen product has
molar impurity fraction $1-x_{\mathrm{N_2}}$,
and molar flow rate $\dot n_{\mathrm{prod}}$.
The control degrees of freedom are the feed air flow rate $\dot n_{\mathrm{mac}}$, 
the stream split fraction to the primary turbine $\xi_{\mathrm{tur}}$,
the reflux fraction at the column top $\xi_{\mathrm{reflux}}$,
and the liquid drain from the reboiler $\dot n_{\mathrm{drain}}$.
The process is equipped with a stabilizing P-controller manipulating
$\dot n_{\mathrm{drain}}$ to counteract the integrating response of the liquid reboiler inventory $M_{\mathrm{r}}$ to the other inputs.
We use a detailed dynamic process model, referred to as digital twin, taken from our previous work \citep{CaspariWave}.
In particular, we formulate mass and energy balances for all process units, including tray-by-tray modeling of the column and finite volume discretization of PHX1 and PHX2.
We complement these equations by thermodynamic correlations, including Margules activity model for equilibrium computations.

The digital twin \citep{CaspariWave} is implemented in the modeling language Modelica and has 118 differential and 2675 algebraic equations.
We assume that the model captures the process response sufficiently accurately,
and we use it as a starting point for model reduction as well as a plant representative in closed-loop operation. 
The model is an index-one nonlinear semi-explicit
differential-algebraic equation system (DAE), having a unique and smooth solution to resemble the behavior of Eq.~(\ref{eqn:outputs}).
All computations run on a terminal server with
Intel XEON E5-2630 v2 CPU at 2.6 GHz and 128 GB RAM.

\begin{tablehere}
	\caption{Parameters of data sampling.} 
	\vspace{2ex}
	\label{tab:sampling}
	\begin{footnotesize}
		\begin{tabular}{llll}
			\toprule
			\textbf{Variable} & \textbf{Symbol} & \textbf{Value} & \textbf{Unit}\\
			\midrule
			Sampling time & $\Delta t_s$ & 5 &\unit{min}\\
			No.~diff.~states & $n_x$ & 118 & $-$\\
			No.~outputs & $n_y$ & 3 & $-$\\
			Dynamic traj.~length & $t_{f,d}$ & 1000 & h\\
			Stationary traj.~length & $t_{f,s}$ & 1000 & h\\
			Air feed rate  & $\dot n_{\mathrm{mac}}$& $[28,\,52]$ & $ \unit{mol\,s^{-1}}$\\
			Turbine split  & $\xi_{\mathrm{tur}}$& $[0.88, 1.0]$ & $-$\\
			Reflux ratio & $\xi_{\mathrm{reflux}}$ &$[0.5, 0.55]$ & $-$ \\
			Reboiler setpoint & $M_{\mathrm{r,sp}}$ & $[19, 41]$ & $\unit{kmol}$ \\
			P-controller gain & $K_p$ & 5 & $\unit{mol\,s^{-1}\,kmol^{-1}}$ \\
			\bottomrule
			\vspace{-5ex}
		\end{tabular}
	\end{footnotesize}
	\vspace{2ex}
\end{tablehere}

\subsection{Data sampling}
The data sampling and model training procedure is similar to our previous work \citep{Schulze.2022a}.
We obtain the training data set by simulating the full-order digital twin subject to a series of input steps comprising 800 random combinations of all control degrees of freedoms.
For the purpose of data sampling, we retain the P-controller to ensure asymptotic stability and vary the inventory setpoint $M_{\mathrm{r,sp}}$.
However, we record the (biased) manipulation of $\dot n_{\mathrm{drain}}$ to enable a Koopman model without P-controller, deciding against training a model of the process with base-layer control here.
The duration of all input steps varies between 0.5 and \unit[2]{h} (\unit[1000]{h} in total) to mix excitation of varying frequency.
Moreover, we add 500 steady-state trajectories of \unit[2]{h} length
(\unit[1000]{h} in total), to ensure stability properties and small steady-state offset of the trained model.

Table \ref{tab:sampling} collects the information about the sampling.
The input ranges are chosen slightly larger than the intended use in order to reduce boundary phenomena.
For the simulation of the digital twin, we specify relative and absolute tolerances of $10^{-5}$ and $10^{-8}$, respectively.
We collect the snapshots as tuples of states, outputs, and controls.
The three algebraic outputs $\bm y$ are the product impurity, production rate, and driving temperature difference in the IRC.

\subsection{Model architecture and training}
To identify suitable hyperparameters,
we perform a systematic parameter study varying the autoencoder morphology (number of neurons and hidden layers) and the latent Koopman state dimension.
Therein, we find that $n_z=30$ and symmetric encoding and decoding with two hidden layers providing a constant relative decrease in the number of neurons, i.e., (76, 48) and (48, 76), respectively, are a suitable choice.
Both encoder and decoder networks use tanh activation and linear output layers.
Since the training data do not indicate oscillatory behavior, we preset a diagonal structure of $A$,
confirmed by a training using full $A$.
We use projection constraints to bound all elements of $A$ to $a_{ii} \geq 0$.

We log-transform all molar fractions and scale all variables between zero and one.
The trajectory data set is created by sliding along the recorded data in a moving horizon fashion, stopping
every 5 sampling instants and copying $s=24$ consecutive snapshots.
We group random sets of 32 trajectories as mini-batches and divide the batched data into \unit[80]{\%} training and \unit[20]{\%} validation data.
Model training is performed for $20\,000$ epochs using the optimizer Adam.
After the training, we retrieve the weights with smallest validation loss.

\begin{figure*}
	\centering
	\includegraphics[width=0.93\linewidth]{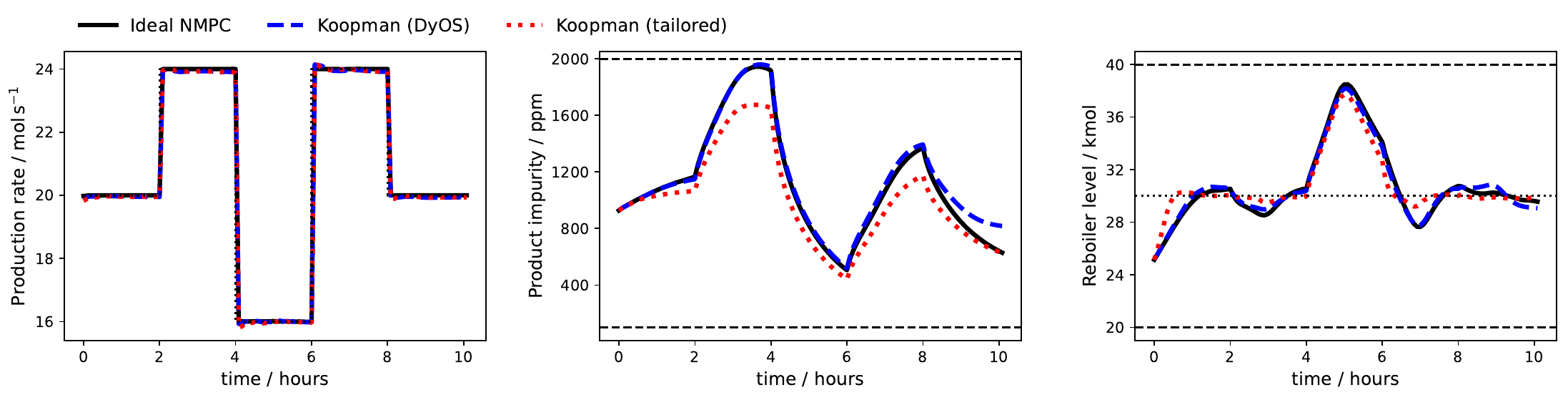}
	\vspace{-2ex}
	\caption{\label{fig:results} Closed-loop results (process response) of the controlled variables in the NMPC case study.}
	\vspace{-1ex}
\end{figure*}

\subsection{Model testing}
We perform an open-loop test of the reduced Koopman model in an independent test scenario in which the model is subject to two random steps of all inputs.
Fig.~\ref{fig:validation} compares the open-loop prediction of the target quantities (production rate, product impurity, and reboiler hold-up) by the Koopman model to trajectories generated using the digital twin.
The graphs show both a single multi-step simulation sweep over the entire horizon and a series of single-step predictions initialized along the reference trajectory.
The single-step predictions are most important for short-term tracking and stabilization, whereas the multi-step prediction reflects the accuracy of the model to open-loop predict long-term trends.
Despite the high degree of reduction, the Koopman model captures both short and long-term trends accurately.
At high product impurities, we observe an opposite gain in a number of single-step predictions, 
which will be the subject of future improvements.
However, this artifact did not affect the NMPC performance in the control study below.

\subsection{Control case study}
Since we target demand side management based on frequent load changes \citep{Pattison.2016b}, the control task is tracking of a series of instantaneous changes in the production rate,
while maintaining product quality and stabilizing the liquid reboiler level $M_{\mathrm{r}}$.
We use the digital twin as plant representation in closed-loop operation and leave the effect of mismatch between digital twin and plant for future work.
We formulate the NMPC tracking stage cost as:
\begin{equation}
\vspace{-2ex}
\ell := w_1 \cdot (\dot n_{\mathrm{prod}} - \dot n_{\mathrm{prod,sp}})^2 
+ w_2 \cdot (M_{\mathrm{r}} - M_{\mathrm{r,sp}})^2
\,.
\end{equation}
Table \ref{tab:tuning} summarizes the NMPC tuning.

\begin{tablehere}
	\caption{NMPC tuning.} 
	\vspace{2ex}
	\label{tab:tuning}
	\begin{footnotesize}
		\begin{tabular}{llll}
			\toprule
			\textbf{Variable} & \textbf{Symbol} & \textbf{Value} & \textbf{Unit}\\
			\midrule
			Sampling time & $\Delta t_s$ & 5 &\unit{min}\\
			Control horizon & $T_c$ & 120 & \unit{min}\\
			Tuning weight 1 & $w_1$ & 1.0 & $\unit{s^{2}\,mol^{-2}}$ \\
			Tuning weight 2 & $w_2$ & 0.0005 &$ \unit{kmol^{-2}}$ \\
			Level setpoint (fixed) & $M_{\mathrm{r,sp}}$ & 3.0 & $\unit{kmol}$ \\
			Level constraint & $M_{\mathrm{r}}$ & [20, 40] & $\unit{kmol}$ \\
			Impurity constraint & $1-x_{\mathrm{N_2}}$ & [100, 2000] & $\unit{ppm}$ \\
			Air feed rate  & $\dot n_{\mathrm{mac}}$& $[30,\,50]$ & $ \unit{mol\,s^{-1}}$\\
			Turbine split  & $\xi_{\mathrm{tur}}$& $[0.9, 1.0]$ & $-$\\
			Reflux ratio & $\xi_{\mathrm{reflux}}$ &$[0.51, 0.54]$ & $-$ \\
			Reboiler drain & $\dot n_{\mathrm{drain}}$ & [0, 1.0]& $\unit{mol\,s^{-1}}$ \\
			\multicolumn{2}{l}{Integrator tolerances (\textit{only DyOS})} & $10^{-6}$ & $-$ \\
			Optimality tolerance && $10^{-5}$ & $-$ \\ 
			Feasibility tolerance &&$10^{-3}$ & $-$ \\
			\bottomrule
		\end{tabular}
	\end{footnotesize}
\vspace{3ex}
\end{tablehere}

We scale all terms in Eq.~(\ref{eqn:nmpc}) and warm-start all optimizations.
The NMPC does not anticipate the setpoint changes, i.e.,
we expect a considerable CPU effort in re-optimization at setpoint updates.
We apply full state feedback to exclude state estimation errors and focus on the model reduction.

We compare the tailored Koopman NMPC implementation to an ideal NMPC optimizing the full-order control model using a general-purpose DAE optimizer (DyOS).
In addition, we compare to Koopman NMPC implemented with DyOS, i.e., an NMPC implementation similar to our previous work \citep{Schulze.2022b}.
These two benchmarks enable us to evaluate two effects separately: 1) The model reduction, 
i.e., benefit from full-order vs.~reduced Koopman using the same optimization platform, 
and 2) the benefit from a tailored optimization exploiting the specific problem structure.
To enable 1), we transform the Koopman model to continuous-time form using MATLAB 2019a and translate it to the Modelica language.
In all cases, we neglect the closed-loop effect of CPU delays on the controlled process.

Fig.~\ref{fig:results} shows the closed-loop results of the case study, i.e., the process response to the control action.
All controllers accomplish fast and precise setpoint tracking of the primary target (Fig.~\ref{fig:results}a)
and feasible operation of the ASU (Figs.~\ref{fig:results}b+c),
consistent with the excellent open-loop predictions by the reduced models (Fig.~\ref{fig:validation}).
While benchmark and Koopman NMPC using the same optimization platform (DyOS) yield almost indistinguishable closed-loop trajectories, the tailored Koopman implementation exhibits slight deviations from the other trajectories, most pronounced for the impurity.
However, these deviations are limited and not associated with a noticeable tracking or feasibility loss.
We attribute the deviations to numerical effects in the sensitivity computations and the respective local optimization algorithms.

\begin{tablehere}
	\caption{CPU effort reduction achieved by the proposed Koopman NMPC implementation.} 
	\vspace{2ex}
	\label{tab:cpu}
	\begin{footnotesize}
		\begin{tabular}{lccc}
			\toprule
			\textbf{NMPC} & \textbf{$\varnothing$ CPU time} & \textbf{$\varnothing$ Red.} & \textbf{Max.~CPU time} \\
			\midrule
			Ideal & \unit[481]{s} & -- & \unit[3358]{s} \\ 
			Koopman (DyOS) & \unit[47]{s} & \textbf{\unit[90]{\%}} & \unit[330]{s} \\ 
			Koopman (tailored) & \unit[9]{s} & \textbf{\unit[98]{\%}} & \unit[48]{s} \\
			\bottomrule
		\end{tabular}
	\end{footnotesize}
	\vspace{1ex}
\end{tablehere}

We compare the CPU effort of solving the NMPC implementations and assess the real-time capability in 
Table \ref{tab:cpu}.
The computational effort of ideal NMPC vastly exceeds the sampling time of \unit[300]{s} and thereby introduces severe control delay.
In contrast, employing the reduced Koopman model in the same NMPC framework generates a speed-up of factor 10, resulting in average CPU cost below the sampling time. 
However, the maximum CPU effort still lies beyond the sampling rate and the average computational delay is noticeable.
Finally, the tailored NMPC facilitates most efficient optimization at an additional speed-up of over factor of 5, yielding an average CPU time decrease by \unit[98]{\%}.
The CPU effort lies well below the sampling time and is of acceptable magnitude.
Further, this speed-up is slightly improved compared to previous works applying classical model-based reduction methods to the same ASU \citep{Schafer.2019b, CaspariWave}, where an average CPU reduction by \unit[95]{\%} was reported.
Compared to these methods, our approach involves considerably less modeling effort, as the full-order process model is not modified in the reduction process and our deep learning framework is automated.
\vspace{-2ex}

\section{Conclusions}
We apply Koopman theory for data-driven model reduction and real-time NMPC of a chemical process, specifically an ASU.
The reduced model is trained using deep learning and consists of an autoencoder and linear latent dynamics, overall referred to as Wiener-type Koopman model.
The data-driven nature of our approach greatly reduces the required process knowledge and enables an automated reduction procedure.
In addition, we present an NMPC implementation
tailored to the block structure of the reduced models.

Despite the high degree of reduction, the low-order models are accurate and enable precise control of the ASU at a reduction of CPU costs by \unit[98]{\%}.
If required, a further speed-up could be achieved by 
applying a more efficient NLP solver,
stronger reduction, or by
increasing the tolerances at the cost of tracking performance.
Additionally, employing decomposition strategies for further exploitation of the graph structure \citep{Jalving.2019} may enable faster optimization.

In this study, we did not consider state estimation to exclude the effect of estimation errors on closed-loop performance, thereby allowing an isolated study of the reduction approach.
However, we have recently proposed a strategy to train state estimation into the data-driven Koopman model structure \citep{Schulze.2022b}.
Future work will apply this integrated approach in extended studies.
Moreover, we will compare against conventional controller types.
Finally, we have assumed that the provided digital twin captures the real plant exactly. 
To account for plant-model mismatch, future work will investigate methods for online model improvement in closed-loop operation.

\section{Acknowledgment}
We gratefully acknowledge the financial support of the Kopernikus project SynErgie 2 by the Federal Ministry of Education and Research (BMBF) and the project supervision by the project management organization 
Projekttr\"ager J\"ulich, as well as from the Helmholtz Association of German Research Centers as part of the Helmholtz School for Data Science in Life, Earth and Energy (HDS-LEE).

\section{References}
\def\refname{}
\def\bibsection{}

\bibliographystyle{chicago}

\begin{small}

\end{small}

\end{document}